\documentclass[manuscript,endfloat]{geophysics}
\usepackage{graphicx}
\usepackage{subfigure}
\usepackage[english]{babel}
\usepackage{amsmath, amsthm, amssymb}
\usepackage{mathtools}
\usepackage{geometry} 
\usepackage{setspace}
\usepackage{siunitx}
\usepackage[dvipsnames]{xcolor}
\begin{document}

\title{Enhanced prediction accuracy with uncertainty quantification in monitoring ${\mathrm{CO_2}}$ sequestration using convolutional neural networks}

\renewcommand{\thefootnote}{\fnsymbol{footnote}} 

\address{
\footnotemark[1]Theoretical Division, Los Alamos National Laboratory, Los Alamos, 87545, USA \\
\footnotemark[2]Center for Wave Phenomena, Colorado School of Mines, Golden, 80401, USA \\
\footnotemark[3]Computational Mathematics, Science and Engineering, Michigan State University, East Lansing, 48824, USA\\
\footnotemark[4]Earth and Environmental Sciences Division, Los Alamos National Laboratory, Los Alamos, 87545, USA
}

\author{Yanhua Liu\footnotemark[1] \footnotemark[2], Xitong Zhang\footnotemark[1]\footnotemark[3], Ilya Tsvankin\footnotemark[2], and Youzuo Lin\footnotemark[4]}



\maketitle



\begin{abstract}
Monitoring changes inside a reservoir in real time is crucial for the success of ${\mathrm{CO_2}}$ injection and long-term storage. Machine learning~(ML) is well-suited for real-time ${\mathrm{CO_2}}$ monitoring because of its computational efficiency. However, most existing applications of ML yield only one prediction~(i.e., the expectation) for a given input, which may not properly reflect the distribution of the testing data, if it has a shift with respect to that of the training data. The Simultaneous Quantile Regression~(SQR) method can estimate the entire conditional distribution of the target variable of a neural network via pinball loss. Here, we incorporate this technique into seismic inversion for purposes of ${\mathrm{CO_2}}$ monitoring. The uncertainty map is then calculated pixel by pixel from a particular prediction interval around the median. We also propose a novel data-augmentation method by sampling the uncertainty to further improve prediction accuracy. The developed methodology is tested on synthetic Kimberlina data, which are created by the Department of Energy and based on a ${\mathrm{CO_2}}$ capture and sequestration~(CCS) project in California. The results prove that the proposed network can estimate the subsurface velocity rapidly and with sufficient resolution. Furthermore, the computed uncertainty quantifies the prediction accuracy. The method remains robust even if the testing data are distorted due to problems in the field data acquisition. Another test demonstrates the effectiveness of the developed data-augmentation method in increasing the spatial resolution of the estimated velocity field and in reducing the prediction error.
\end{abstract}

%
%

\section{Introduction}
Since the rise of industrialization in the 18th century, human activities have increased the volume of atmospheric ${\mathrm{CO_2}}$ by 50\%, thus increasing the global temperature. Capturing industrial ${\mathrm{CO_2}}$ at its various sources and injecting it into geologic formations for long-term storage (sequestration) is one of the most promising methods to combat global warming. 

Seismic data provide valuable information about the subsurface, and can be used to monitor ${\mathrm{CO_2}}$ injection. Both laboratory and field data confirm that P-wave velocity decreases with ${\mathrm{CO_2}}$ saturation, especially for relatively low saturation levels~\citep{Kim2010}. Seismic images can also help delineate ${\mathrm{CO_2}}$ plumes inside the reservoir. Therefore, seismic data can be used to monitor CO${_2}$ injection and storage in subsurface structures and, potentially, detect small leakages of ${\mathrm{CO_2}}$~\citep{Lumley2010,Furre2017, Pevzner2017}. 

Data-driven neural networks have been applied to monitor the ${\mathrm{CO_2}}$ movement using time-lapse seismic data. For example, \cite{Li2021-co2} develop a fully-connected neural network to map the relationship between time-lapse seismic data and the velocity changes caused by the injected $\mathrm{CO_2}$. \cite{Feng2021-CO2}  propose spatio-temporal neural-network-based models with long short-term memory (LSTM) structure to monitor and forecast the ${\mathrm{CO_2}}$ storage at Sleipner field in the North Sea. \cite{Liu2022-Data} use a hybrid time-lapse strategy that combines physics-based FWI and data-driven neural network to monitor the ${\mathrm{CO_2}}$ movement in the reservoir. The main advantage of data-driven inversion is its efficiency during the application stage~(after training), which is essential in real-time monitoring of ${\mathrm{CO_2}}$ sequestration~\citep{ZhangZP2020}. An extensive overview of data-driven seismic inversion methods can be found in \cite{Physics-2022-Lin}.

However, deterministic neural networks assume that the mapping learned by the network is accurate, which is not always the case, and produce only the learned output for a given input. On the other hand, uncertainty quantification~(UQ) not only describes predictive distributions over outputs for given inputs, but also indicates whether the model is confident about the prediction. The uncertainty can be divided into two categories based on its sources: epistemic~(model) and aleatoric~(data) uncertainty~\citep{Natasa2019}. Epistemic uncertainty is produced by the neural network itself: its architecture, training procedures, the number of samples, etc. It can be mitigated by collecting more representative training data, which helps improve testing performance~\citep{ren2021survey}. Data uncertainty describes the variance of the conditional distribution of a prediction for given input features. In contrast to epistemic uncertainty, data uncertainty cannot be reduced by modifying model architecture, training algorithms, or collecting more data under the same experimental conditions because the noise distribution in seismic data cannot be considered constant.

Uncertainty evaluation has been used in seismic monitoring of ${\mathrm{CO_2}}$ injection to quantify prediction accuracy. \cite{chen2018geologic} propose to monitor ${\mathrm{CO_2}}$ leakage by multivariate adaptive regression splines and to measure the prior and posterior uncertainty using the percentile estimation from Monte-Carlo simulations. \cite{tang2022deep} develop a 3D recurrent R-U-Net surrogate model to predict ${\mathrm{CO_2}}$ saturation of a synthetic sequestration region and quantify the uncertainty based on rejection sampling using the CNN~(convolutional neural network)-PCA (principle-component analysis) model. \cite{Um2022} introduce a U-Net network to estimate ${\mathrm{CO_2}}$ saturation and model uncertainty with two UQ methods (i.e., the Monte Carlo dropout method and a bootstrap aggregating method).

However, the uncertainty quantification methods mentioned above either require an extra sampling step during testing or use the predictive variance/disagreement to represent the uncertainty. The sampling step can be time-consuming, if physics simulation is involved. Moreover, using just one variable~(e.g., variance) as the uncertainty indicator assumes the uncertainty to be symmetric around the estimated prediction, which is not always true. For example, the uncertainty becomes asymmetric, if the noise inherent in the input data is sampled from a skew-normal distribution.

The Simultaneous Quantile Regression~(SQR) is proposed by \cite{Natasa2019} to quantify the uncertainty of a 1D regression problem. SQR is designed to train a model that makes predictions at different quantile levels utilizing pinball loss. The uncertainty map is computed by subtracting the predictions at two selected quantile levels around the predicted median values. Hence, SQR does not need extra sampling for uncertainty estimation and can produce asymmetric confidence intervals. 

Here, we incorporate SQR into a CNN designed to estimate the uncertainty of the velocity distribution estimated from seismic data. To improve the prediction accuracy, we propose a novel data-augmentation method that operates with the prediction and calculated uncertainty. We begin by discussing the methodology of SQR and the architecture of the proposed network~(InvNet\_UQ). Then InvNet\_UQ is tested on the Kimberlina model based on a ${\mathrm{CO_2}}$ injection project in California. The performance of InvNet\_UQ is evaluated by comparing the predictions with those of a deterministic neural network from both the perfect and distorted testing data. The robustness of the computed uncertainty is verified by its comparison with the prediction error. Finally, the effectiveness of the data-augmentation method is evaluated by comparing the error of the predicted velocity model before and after applying the augmentation.

\section{Method}
\begin{figure}
    \centering
    \includegraphics[trim=0cm 6cm 2cm 0cm,clip,scale=0.22,width=0.9\textwidth]{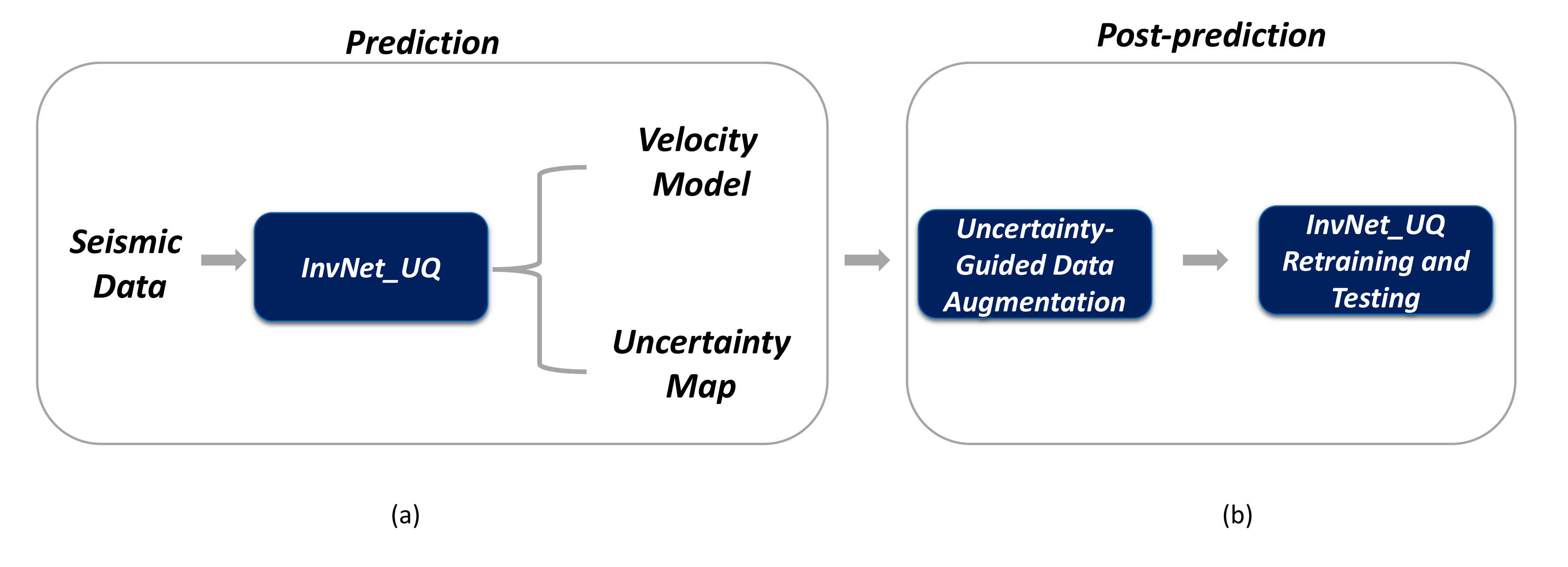}
    \caption{Flowchart of the method. (a) InvNet$\_$UQ is trained and tested to estimate the velocity and the corresponding uncertainty. (b) The uncertainty-guided data-augmentation method is applied to the predicted velocity models; then the network is retrained with the augmented data.}
    \label{fig:framework}
\end{figure}

Figure~\ref{fig:framework} illustrates the developed workflow using the proposed probabilistic convolutional neural network~(InvNet\_UQ). First, InvNet\_UQ is trained and applied to predict the velocity distributions at all quantile levels from the testing data. Meanwhile, the uncertainty map is calculated from two predictions at different quantile levels. Next, uncertainty-guided data augmentation is applied to all predicted velocity models from the testing data. Finally, the augmented data are added to the existing training data for retraining the network. 
\subsection{Simultaneous Quantile Regression} 
Uncertainty estimation is essential for seismic inversion for two reasons. First, seismic data contain different kinds of noise caused by the recording equipment, ambient disturbances, etc. Second, inversion is often ill-posed, which means that the solution can be nonunique or correspond to a local minimum of the objective function. Here, we estimate the uncertainty using the probabilistic neural based on the SQR proposed by \cite{Natasa2019}. 

To solve the regression problem, we implement a neural network as a function ${\hat{y} = \hat{f}_{\tau}(x)}$ to approximate the relationship between the input ${x}$ and output ${y}$. To identify the network parameters, we minimize the mean-square error~(MSE) between the prediction~(${\hat{y}}$) and actual~(${y}$) values:
\begin{equation}
    MSE = \frac{1}{n} \sum_{i=1}^{n} (y_i - \hat{y}_i)^2 = E(y - \hat{y})^2.
\end{equation}
Therefore, the prediction (${\hat{y}}$) represents the conditional mean~(expectation) of the prediction for a given input.

However, the expectation alone does not accurately reflect the data distribution. Therefore, quantile regression is proposed to analyze the prediction in a specific quantile (${\tau}$). One way to build such a model approximating the conditional quantile distribution function ${y = F^{-1}(\tau | X = x)}$ is to minimize the pinball loss (${l_{\tau}(y, \hat{y})}$):
\begin{equation}
    l_{\tau}(y, \hat{y}) = 
    \begin{cases}
        \tau (y - \hat{y}) & \text{if ${y-\hat{y} \geq 0}$},\\
        (1-\tau) (\hat{y} - y) & \text{else}.
    \end{cases}
\end{equation}
Indeed,
\begin{equation}
    E[l_{\tau}(y, \hat{y})] = (\tau - 1) \int_{-\infty}^{\hat{y}} (y - \hat{y}) dF(y) + \tau \int_{\hat{y}}^{\infty} (y - \hat{y}) dF(y),
\end{equation}
where ${F(y) = P(Y \leq y)}$ is the strictly monotonic cumulative distribution function of the target variable ${Y}$ taking real values ${y}$. Consequently, ${F^{-1}(\tau) = inf\{y: F(y) \geq \tau\}}$ denotes the quantile distribution function of the same variable ${Y}$ for all quantile levels ${0 \leq \tau \leq 1}$. 

With pinball loss, \cite{Natasa2019} propose employing Simultaneous Quantile Regression~(SQR) to estimate all the quantile levels simultaneously by solving the following equation:
\begin{equation}
    \hat{f} \in \text{arg min}_{f} \frac{1}{n} \sum_{i = 1}^{n} E_{\tau \approx U[0,1]} [l_{\tau} (f(x_i, \tau), y_i)],
    \label{eq:UQ_Loss}
\end{equation}
where $(x_i, y_i)$ are identically and independently distributed feature-target pairs drawn from the unknown probability distribution ${P(X, Y)}$. Note that, unlike Bayesian neural networks, ${P(X,Y)}$ is not necessarily a Gaussian distribution. Then the estimated data uncertainty can be computed from the (${1-\alpha}$) prediction interval around the median:
\begin{equation}
    u_a(x^{\star}):=\hat{f}(x^{\star}, 1-\frac{\alpha}{2}) - \hat{f}(x^{\star}, \frac{\alpha}{2}),
    \label{eq:UQ}
\end{equation}
where ${\alpha}$ is the significance level.

\subsection{Network architecture of InvNet\_UQ}
Following \cite{wu-2019-inversionnet}, we use an encoder-decoder-based CNN to approximate the relationship between the input seismic data and the output velocity model. There are seven convolutional blocks with a kernel size of ${3\times3}$ followed by a ${2\times2}$ max pooling. Each block consists of a convolutional layer, batch normalization, and a tanh activation function. The high-dimensional data are reduced to a ${1024\times1\times1}$ array in the latent space. These features are upsampled to the output size by eight decoder blocks containing a deconvolutional layer, batch normalization, and a LeakyRelu activation function. The end of the decoder is a central-cropping layer, which crops the output of the deconvolutional blocks to the desired size.

For the benchmark network~(InversionNet), the loss function is the ${L_1}$-norm of the difference between the prediction and the actual model. The input data size is ${6\times1,000\times200}$, where the first number (${6}$) is the number of channels~(shot gathers). For each shot gather, there are 200 receivers recording for 4~s with a time interval of 4~ms. The output is a velocity model with a size of ${351\times601}$. 

InvNet\_UQ shares the same structure with InversionNet, but its loss function is computed using equation~\ref{eq:UQ_Loss}. To compute the SQR loss, first we randomly select six quantile levels~($\tau$) for each training epoch, and the selected ${\tau}$ is then fixed for every batch in the epoch. Then the computed six quantile losses are averaged to obtain the final loss. To incorporate quantile level into the neural network, every ${\tau}$ is extended to a tensor with the exact size of the seismic data and then added to the input data, so that the number of input channels of InvNet\_UQ is increased to seven. 

\subsection{Uncertainty-guided data augmentation}
Most existing UQ methods produce the variance of the prediction as the uncertainty map and abandon it after evaluating the prediction accuracy. In contrast, we propose a data-augmentation method to demonstrate the potential of the estimated uncertainty in improving the network performance. The idea of the method is to sample the uncertainty map ${u_a(x^{\star})}$ and add it to the predicted velocity ${\hat{f}(x^{\star})}$ to obtain new velocity models ${\hat{y}_{aug}(x^{\star})}$:  
\begin{equation}
    \hat{y}_{aug}(x^{\star}):=\hat{f}(x^{\star}) + w \times u_a(x^{\star}),
    \label{eq:Aug}
\end{equation}
where $w$ is the weight matrix applied to the computed uncertainty. Ideally, one needs to sample the uncertainty of every pixel with every weight in the range from 0 to 1 to find the actual velocity model. However, this procedure is impossible because it requires an infinite number of weight matrices to fully cover the range ${[0, 1]}$. To simplify the sampling process and ensure the effectiveness of the data-augmentation method, the weights should be randomly selected and have a minimum of 0 and maximum of 1.

An acoustic forward-modeling algorithm simulates seismic data for these new velocity models with the same settings used to generate the initial data. Then we add these new seismic-velocity pairs to the existing training data and retrain the network. Finally, the retrained network is applied to the testing data to estimate the velocity distribution. Because the new velocity models are generated using the prediction and the estimated uncertainty, they provide extra information that the model does not learn from the training data set. Moreover, the inclusion of these new models makes the training data more representative of the testing data distribution.  

\section{Results}
\subsection{Kimberlina data set}
The proposed neural network with UQ is applied to monitoring and predicting ${\mathrm{CO_2}}$ migration using the synthetic Kimberlina data set. The Kimberlina reservoir model, generated by several institutions~\citep{Development-2021-Alumbaugh} as part of the U.S. Department of Energy ``SMART Initiative''~\citep{SMART-2019-DOE}, was built to simulate a potential commercial-scale geologic carbon storage in the Southern San Joaquin Basin of California~\citep{Wagoner2009}, 30~km northwest of Bakersfield, CA, USA. The Kimberlina model has been released to the public to evaluate the effectiveness and robustness of different geophysical techniques for monitoring ${\mathrm{CO_2}}$ migration.

\begin{figure}
  \centering
  \subfigure{\includegraphics[trim=0cm 2cm 4cm 0cm,clip,scale=0.22,width=0.95\textwidth]{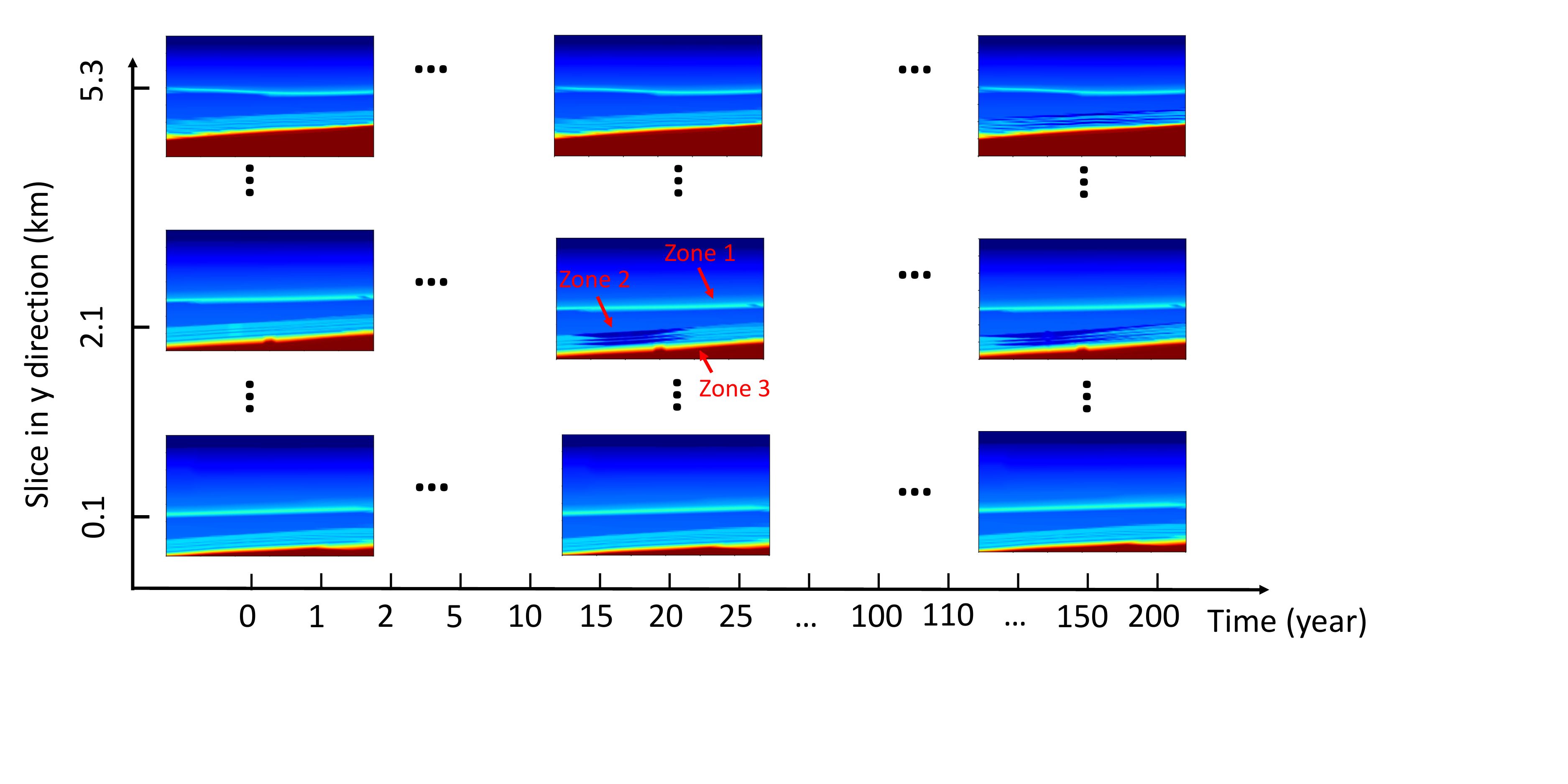}}
  \caption{Visualization of the actual 2D models sliced from the 3D Kimberlina time-lapse velocity model along the ${y}$-direction. Three important geologic structures are marked by red arrows. } 
  \label{fig:geo}
\end{figure}

There is a total of 29 3D time-lapse P-wave velocity models~(601 $\times$ 601 $\times$ 351 grid points) simulated for over 200 years. We slice the 3D velocity model along the ${y}$ axis to obtain the corresponding 2D models. The 2D velocity models from different slices have distinct velocity distributions, while those from the same slice~(Figure~\ref{fig:geo}) share the velocity field outside the reservoir, but inside the reservoir the velocity distribution is changed by the injected ${\mathrm{CO_2}}$. To simulate the seismic data, we place six shots at the surface of the model, which has a grid size of 10 $\times$ 10 $\times$ 10 m. 

The entire Kimberlina data set~(1,537 velocity samples) is randomly divided into training~(80${\%}$) and testing~(20${\%}$) data. InverionNet and InvNet${\_}$UQ are trained with the same data until they converge. Then these networks are applied to the testing data in different scenarios to predict the velocity distribution. To evaluate the predictions, we focus on three critical geologic structures~(Figure~\ref{fig:geo}; pointed by red arrows): the low-velocity layer immediately above the reservoir region~(Zone~1), the three reservoirs~(Zone~2), and the high-velocity dipping layer beneath the reservoirs~(Zone~3).

\subsection{Noise-free data}
\begin{figure}
  \centering
  \subfigure{\includegraphics[trim=0cm 0cm 2cm 0cm,clip,scale=0.22,width=0.95\textwidth]{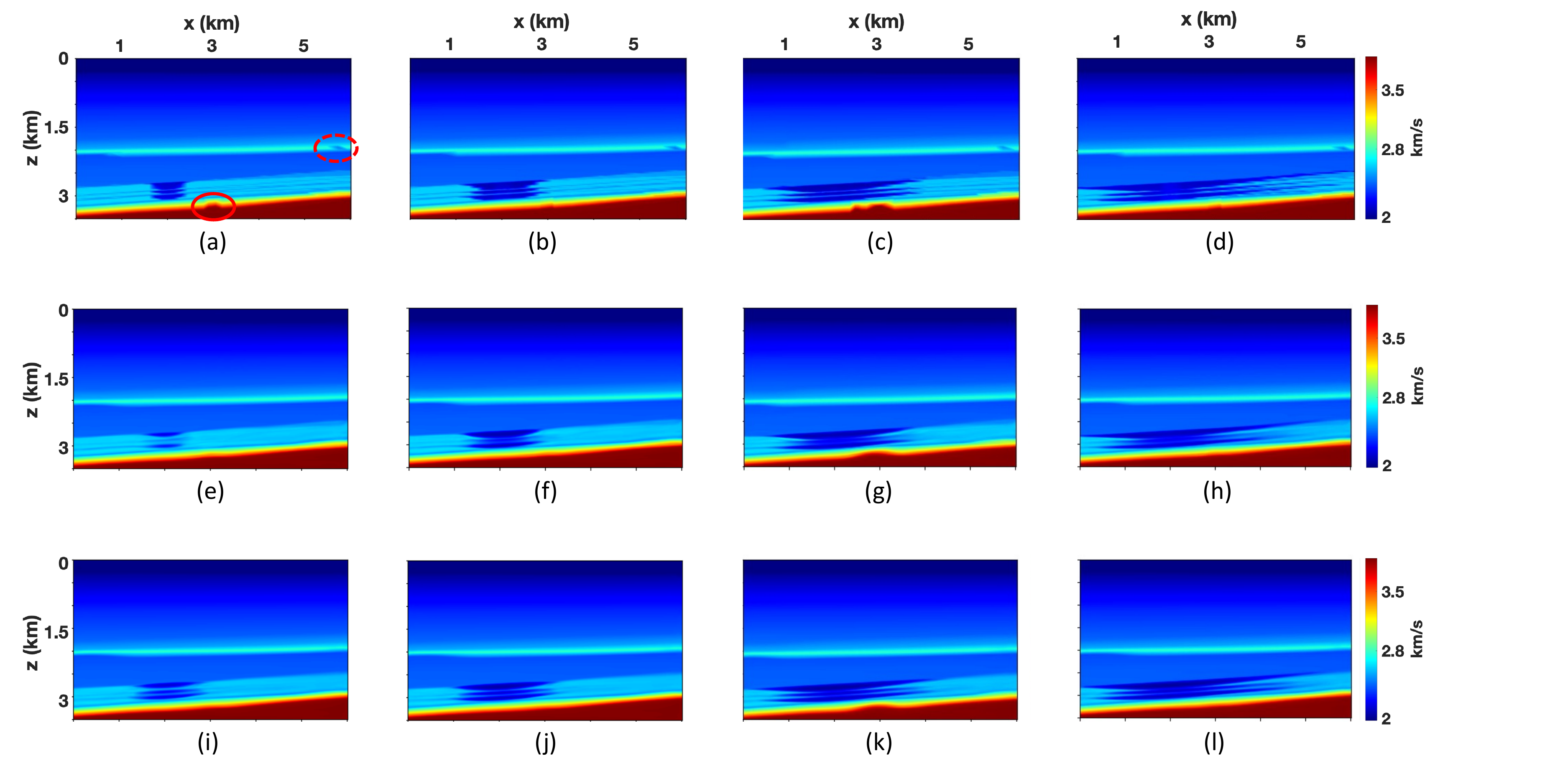}}
  \caption{Actual P-wave velocity in (a) year 1, (b) year 5, (c) year 20, and (d) year 150. The P-wave velocity obtained from noise-free seismic data by (e-h) InversionNet and (i-l) InvNet${\_}$UQ. } 
  \label{fig:clean}
\end{figure}

First, we test the trained networks on the original~(perfect) testing data. The predictions~(Figure~\ref{fig:clean}) show that most features of the velocity model can be predicted by both InversionNet~(Figure~\ref{fig:clean}e-h) and InvNet${\_}$UQ~(Figure~\ref{fig:clean}i-l) with sufficient accuracy, although there exists an area with larger errors. For example, compared with the actual model~(Figure~\ref{fig:clean}), both networks cannot accurately reconstruct the small plumes at year 1~(Figure~\ref{fig:clean}a) because of the lack of training data at the beginning of the injection when the plumes are small. InversionNet tends to underestimate the size of the plumes~(Figure~\ref{fig:clean}e), while InvNet${\_}$UQ is more likely to overestimate them~(Figure~\ref{fig:clean}i).

In addition, both methods could not reconstruct the bumps in the high-velocity dipping layer~(circled by a dashed red line) and the anomaly in the low-velocity horizon above the reservoirs~(circled by a solid red line) because of the lack of training data for these structures. 

\begin{figure}
  \centering
  \subfigure{\includegraphics[trim=0cm 0cm 2cm 0cm,clip,scale=0.22,width=0.95\textwidth]{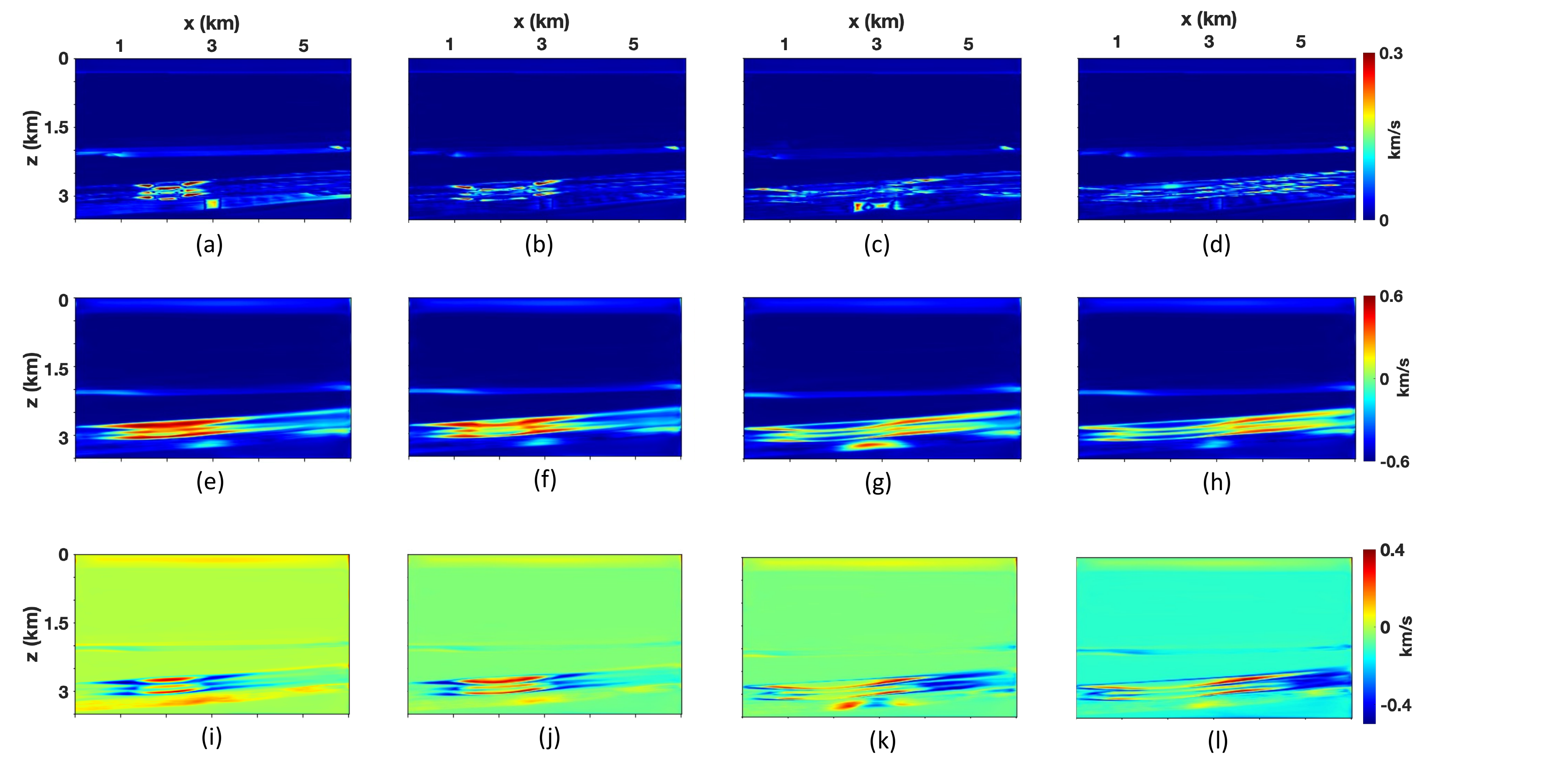}}
  \caption{Difference between the actual velocity models and those predicted by InvNet${\_}$UQ in (a) year 1, (b) year 5, (c) year 20, and (d) year 150. Plots (e-h) are the corresponding uncertainty maps. Plots (i-l) show the asymmetric uncertainty estimated from $\mathrm{(p_{upper}-p_{median})-(p_{median}-p_{lower})}$, where $\mathrm{p_{upper}}$ and $\mathrm{p_{lower}}$ are the predictions at quantiles ${\tau = 0.975}$ and ${\tau = 0.025}$, respectively, and $\mathrm{p_{median}}$ is the median prediction at ${\tau = 0.5}$.} 
  \label{fig:cleanerr}
\end{figure}

For the uncertainty map, the 95${\%}$ prediction interval around the median is calculated from equation~\ref{eq:UQ} by setting the prediction interval ${\alpha=0.05}$. We verify the uncertainty predicted by InvNet${\_}$UQ~(Figures~\ref{fig:cleanerr}e-h) by comparing it with the error map~(Figures~\ref{fig:cleanerr}a-d). High uncertainty is observed around the low-velocity area above the reservoirs, especially near the anomaly~(Zone 1) and the reservoir~(Zone 2), as well as near the small bump in Zone 3, which matches the area with the most significant errors. The difference between the two confidence intervals, calculated from $\mathrm{(p_{upper}-p_{median})-(p_{median}-p_{lower}) \neq 0}$~(Figure~\ref{fig:cleanerr}i-l), demonstrates the usefulness of the asymmetric uncertainty prediction, which cannot be provided by conventional UQ methods.

\begin{figure}
  \centering
  \subfigure{\includegraphics[trim=0cm 5.5cm 20cm 0cm,clip,scale=0.22,width=0.6\textwidth]{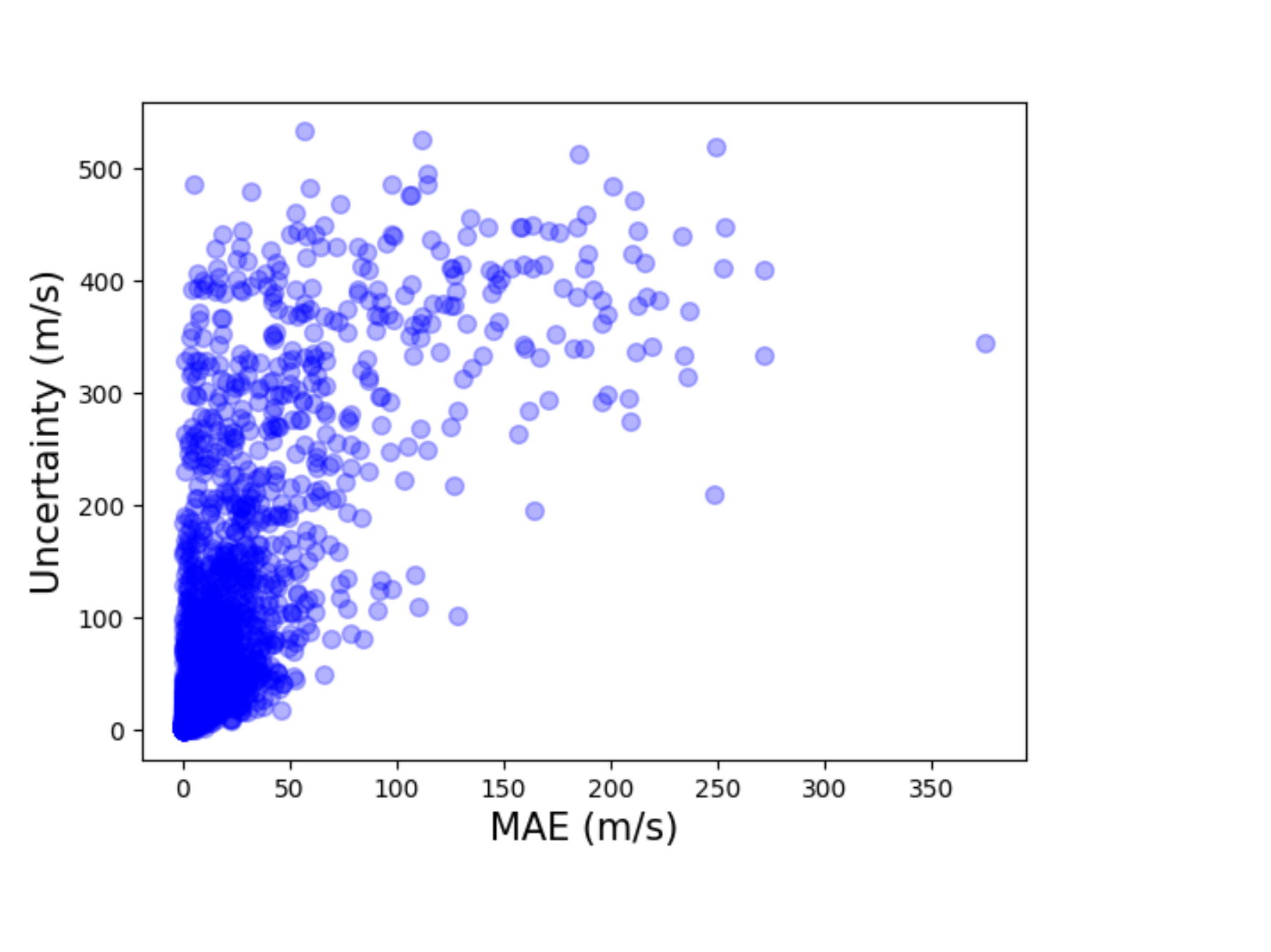}}
  \vskip -1cm
  \caption{Correlation map between the absolute prediction error~(MAE) and the corresponding uncertainty. The Pearson correlation is ${0.727}$ (p-value is \num{1.0e-10}), which shows a strong positive correlation between the absolute prediction error and the uncertainty.} 
  \label{fig:qr-corre}
\end{figure}

We also calculate the Pearson correlation between the absolute prediction error and the uncertainty and present the corresponding scatter plot in Figure~\ref{fig:qr-corre}. The Pearson correlation is $0.727$, and the p-value is \num{1.0e-10}, which shows a strong positive correlation between the two variables. This correlation implies that the uncertainty is an accurate indicator of the prediction confidence, and indicates that InvNet\_UQ can produce an uncertainty map sufficient to facilitate decision-making. 


\subsection{Influence of random noise}
\begin{figure}
  \centering
  \subfigure{\includegraphics[trim=0cm 6cm 0cm 0cm,clip,scale=0.22,width=0.95\textwidth]{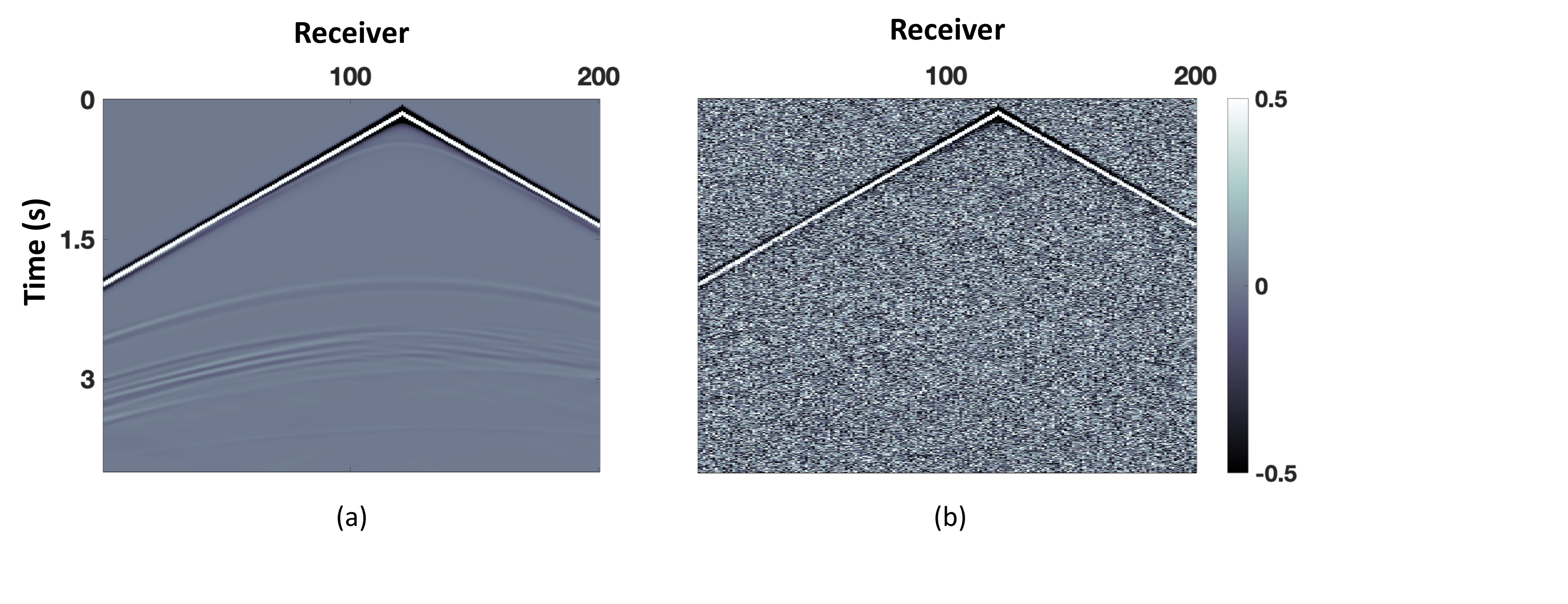}}
   \vskip -10pt
  \caption{Seismic shot gather for the testing velocity model in year 20 (a) without noise, and (b) with Gaussian noise~(the signal-to-noise ratio SNR is 10).} 
  \label{fig:snr10-data}
\end{figure}

Next, the testing data are contaminated with Gaussian noise with the signal-to-noise ratio equal to 10 and 20. The noise is added only to the testing data, whereas the neural networks are still trained on noise-free samples. Here, we show only the prediction results for (SNR=10), which is more realistic for field data. Compared with the clean data~(Figure~\ref{fig:snr10-data}a) where reflection events are apparent, reflections are not clearly visible in the noisy data~(Figure~\ref{fig:snr10-data}b). Note that testing on noisy data is performed without fine-tuning the network. 

\begin{figure}
  \centering
  \subfigure{\includegraphics[trim=0cm 0cm 2cm 0cm,clip,scale=0.22,width=0.95\textwidth]{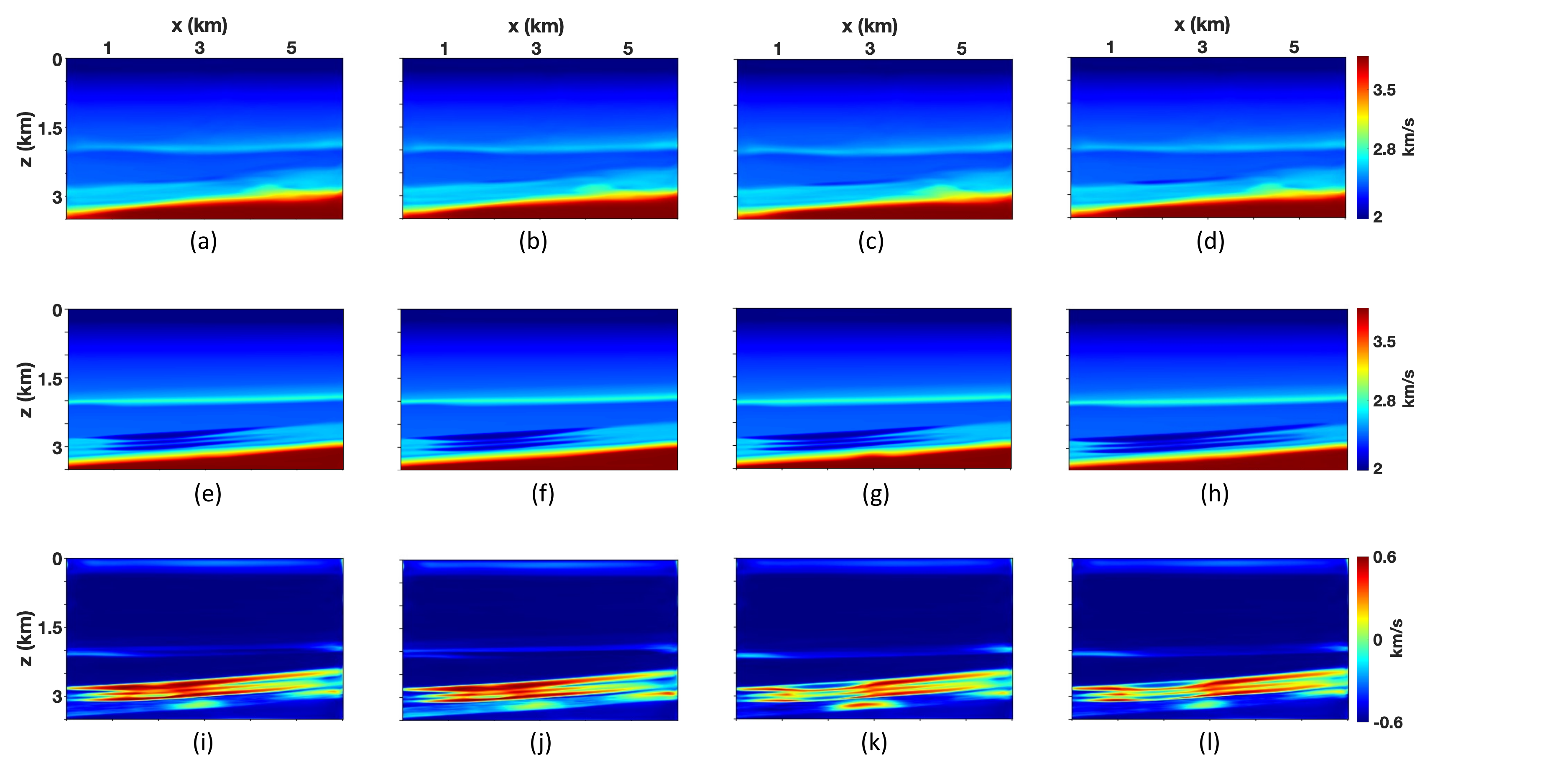}}
  \caption{P-wave velocity obtained from seismic data with SNR=10 by (a-d) InversionNet and (e-h) InvNet${\_}$UQ. Plots (i-l) show the corresponding uncertainty computed by InvNet${\_}$UQ.} 
  \label{fig:snr10}
\end{figure}

Predictably, both networks produce more errors in the velocity field estimated from the noise-contaminated data~(Figure~\ref{fig:snr10}). For the velocity model obtained by InversionNet~(Figure~\ref{fig:snr10}a-d), significant distortions are observed in the low-velocity layer above the ${\mathrm{CO_2}}$ plumes~(Zone~1), in the plumes themselves~(Zone~2), and near the boundary between the reservoir and the deep high-velocity horizons~(Zone~3). The ${\mathrm{CO_2}}$ plumes are barely visible, which would complicate monitoring ${\mathrm{CO_2}}$ injection. 

In contrast, InvNet${\_}$UQ~(Figure~\ref{fig:snr10}e-h) reconstructs Zones~1 and~3 with sufficient resolution. The errors are mainly concentrated near the ${\mathrm{CO_2}}$ plumes in Zone~2, which is consistent with the calculated uncertainty map~(Figure~\ref{fig:snr10}i-l). In addition, the noise does not significantly distort the predictions, which confirms that the proposed network remains robust for at least moderate noise levels. 

\begin{figure}
  \centering
  \subfigure{\includegraphics[trim=0cm 6cm 10cm 0cm,clip,scale=0.22,width=0.6\textwidth]{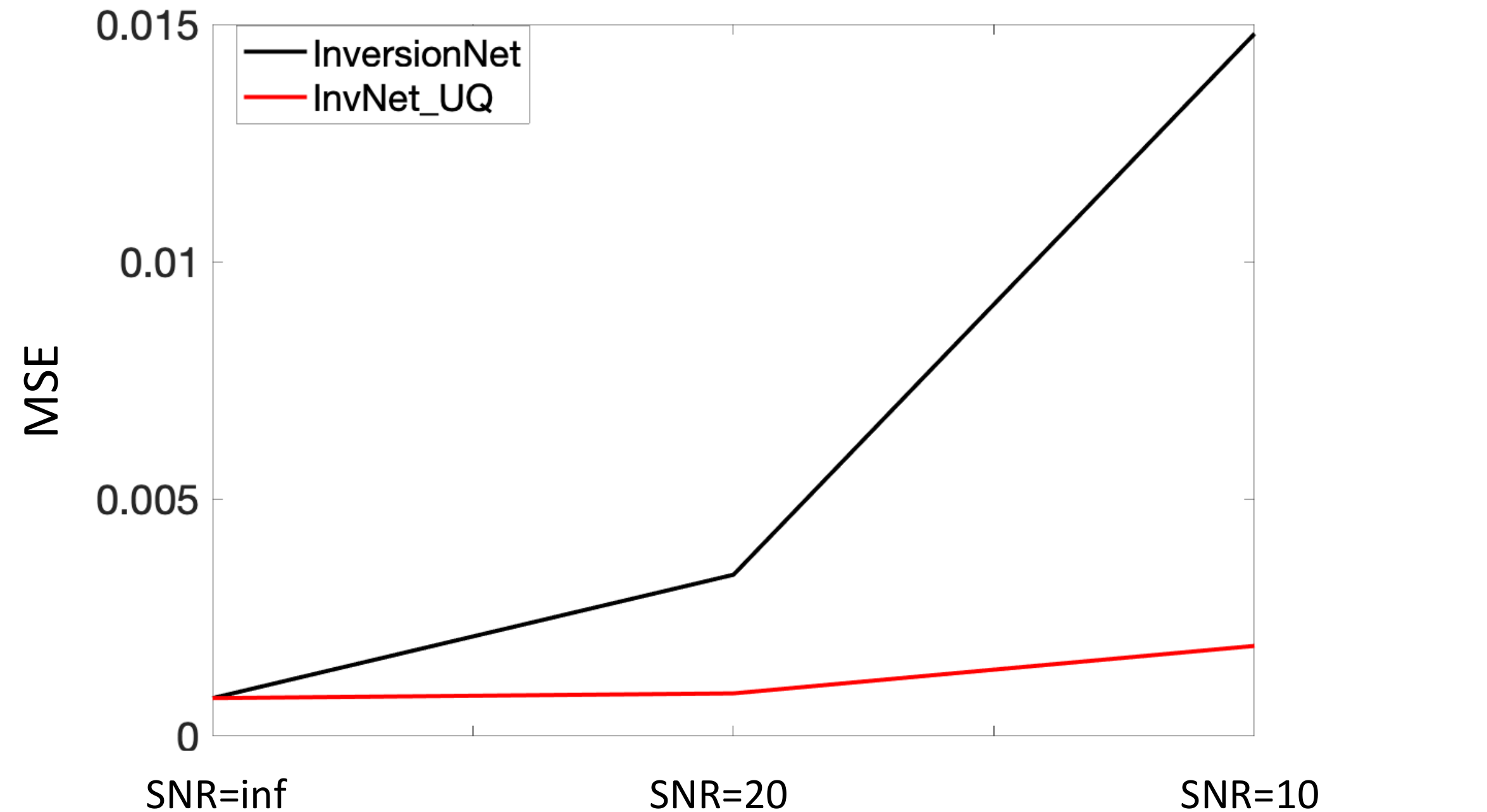}}
  \caption{Comparison of InversionNet and InvNet\_UQ in terms of the mean-square error~(MSE) {vs.} signal-to-noise ratio~(SNR). } 
  \label{fig:mse-snr}
\end{figure}

The MSE~(Figure~\ref{fig:mse-snr}) of the predictions by InversionNet jumps from \num{7.7e-4} for the clean data to \num{1.5e-2} for the noisy data~(SNR=10). In contrast, there is only a slight increase in the MSE of InvNet${\_}$UQ~(i.e., from \num{8.0e-4} to \num{1.9e-3}). Evidently, our method can handle noisy data with a realistic signal-to-noise ratio, and the noise does not distort the uncertainty quantification.

\subsection{Influence of missing traces}
\begin{figure}
  \centering
  \subfigure{\includegraphics[trim=0cm 6cm 0cm 0cm,clip,scale=0.22,width=0.95\textwidth]{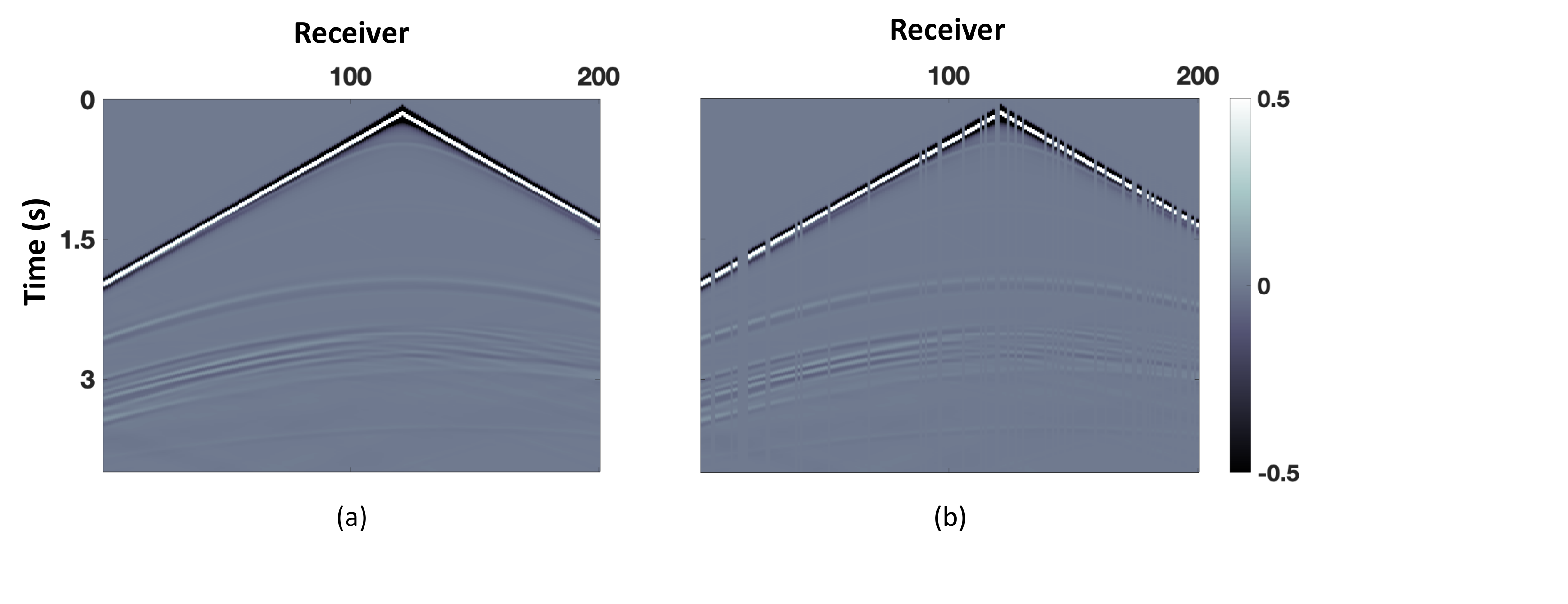}}
   \vskip -10pt
  \caption{Seismic shot gather for the testing velocity model in year 20 (a) without missing data, and (b) with 30${\%}$ of data missing.} 
  \label{fig:missing-data}
\end{figure}

Missing traces are typical in seismic surveys due to unavailable receivers and low SNR on some records. The commonly used remedies are reconstruction of the missing traces using interpolation and compressive sensing. However, accurate reconstruction usually takes a significant amount of time and is computationally costly. Therefore, next we test the effectiveness of InvNet${\_}$UQ in dealing with the missing-data problem by randomly blocking 30${\%}$ traces~(Figures~\ref{fig:missing-data}b) for every testing sample. The networks are trained with the complete seismic data. Hence, testing data that miss traces can be referred to as an ``out-of-distribution" case.

\begin{figure}
  \centering
  \subfigure{\includegraphics[trim=0cm 0cm 0cm 0cm,clip,scale=0.22,width=0.95\textwidth]{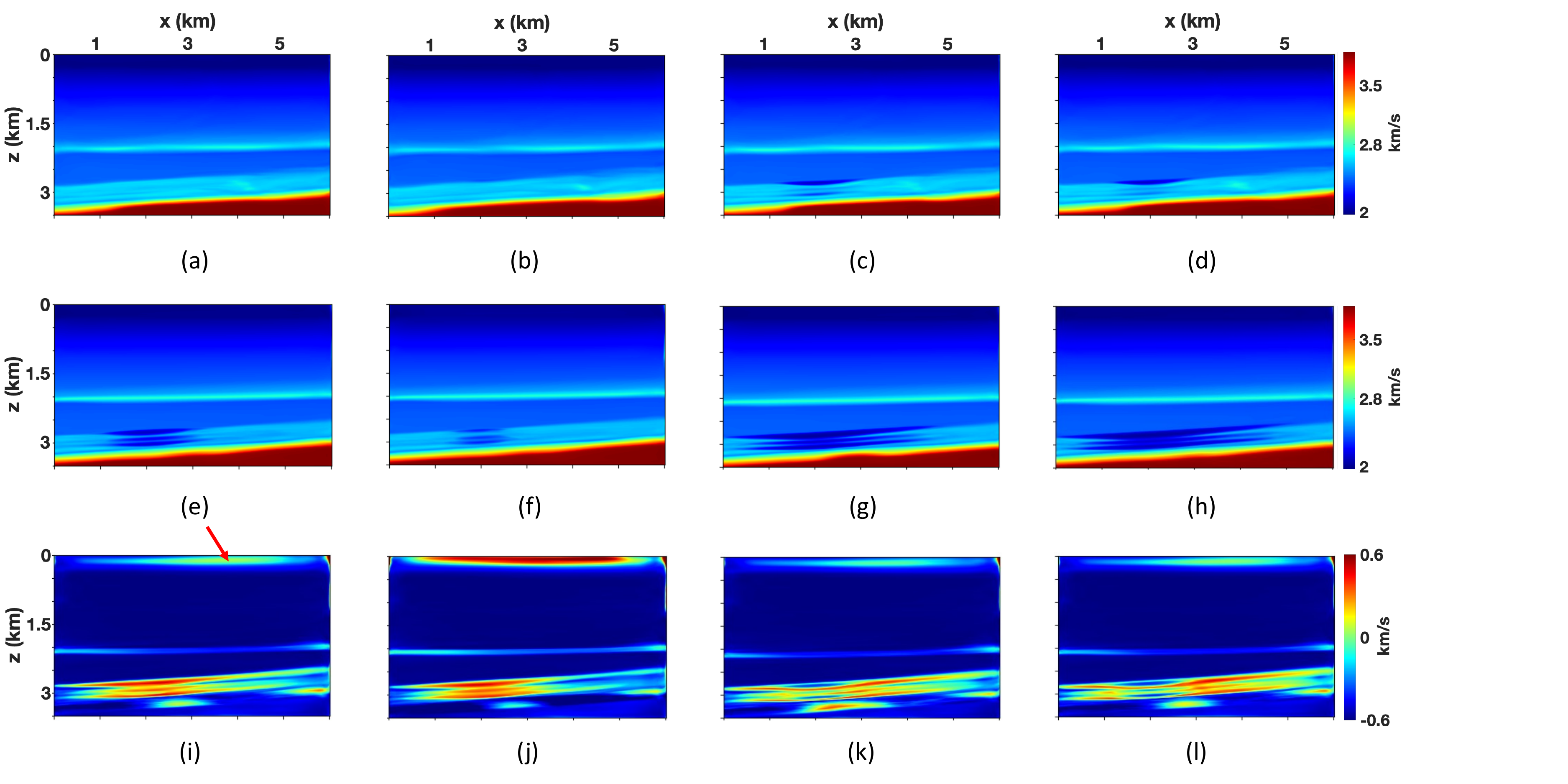}}
  \caption{P-wave velocity obtained from the testing seismic data with 30${\%}$ of the traces missing by (a-d) InversionNet and (e-h) InvNet${\_}$UQ. Plots (i-l) show the corresponding uncertainty from InvNet${\_}$UQ.} 
  \label{fig:missing}
\end{figure}

Compared with the predicted results using the entire data set, both methods produce distortions in the velocity model due to the missing traces. InversionNet~(Figures~\ref{fig:missing}a-d) cannot predict the ${\mathrm{CO_2}}$ plume, while InvNet${\_}$UQ~(Figures~\ref{fig:missing}e-h) still reconstructs these plumes with acceptable resolution. In addition, InvNet${\_}$UQ is able to estimate the bottom right part of the velocity model, which is substantially distorted by InversionNet. Table~\ref{tab:MSE} shows that our method is more robust and reduces MSE by about 30\% compared with InversionNet.

The uncertainty for the upper part of the model~(Figures~\ref{fig:missing}i-l; marked by the red arrow) caused by the missing trace is significantly larger than that for the original data. Similar to the previous tests, the reservoirs generally have a higher uncertainty than other parts of the model. 

\begin{table}[ht]
\centering
\begin{tabular}{|l|l|l|l|}
\hline
Conditions/Methods & InversionNet & InvNet\_UQ & Augmented InvNet\_UQ\\
\hline
Noise-free testing data & \num{7.7e-4} & \num{8.0e-4} & \num{6.0e-4}\\
\hline
Testing data with 30\% missing traces & \num{8.0e-3} & \num{5.6e-3} & \num{9.0e-4}\\
\hline
Testing data with absence of low frequencies & \num{9.3e-4} & \num{8.0e-4} & \num{6.0e-4}\\
\hline
\end{tabular}
\caption{\label{tab:MSE}Comparison of the MSE values for the velocity distribution predicted by InversionNet and InvNet\_UQ applied without data augmentation (columns 2 \& 3) and with data augmentation (column 4).}
\end{table}

\subsection{Influence of absent low-frequency data}
Ultra-low-frequency seismic data are essential for the success of FWI, but they are seldom acquired in the field. Therefore, it is important to test the robustness of our algorithm in the absence of low frequencies~(less than 3~Hz). We apply the Fourier transform to the perfect testing data and remove frequencies below 3~Hz with a Butterworth filter. Note that both neural networks are trained on the perfect training data that include the full frequency range.

\begin{figure}
  \centering
  \subfigure{\includegraphics[trim=0cm 8cm 0cm 0cm,clip,scale=0.22,width=0.95\textwidth]{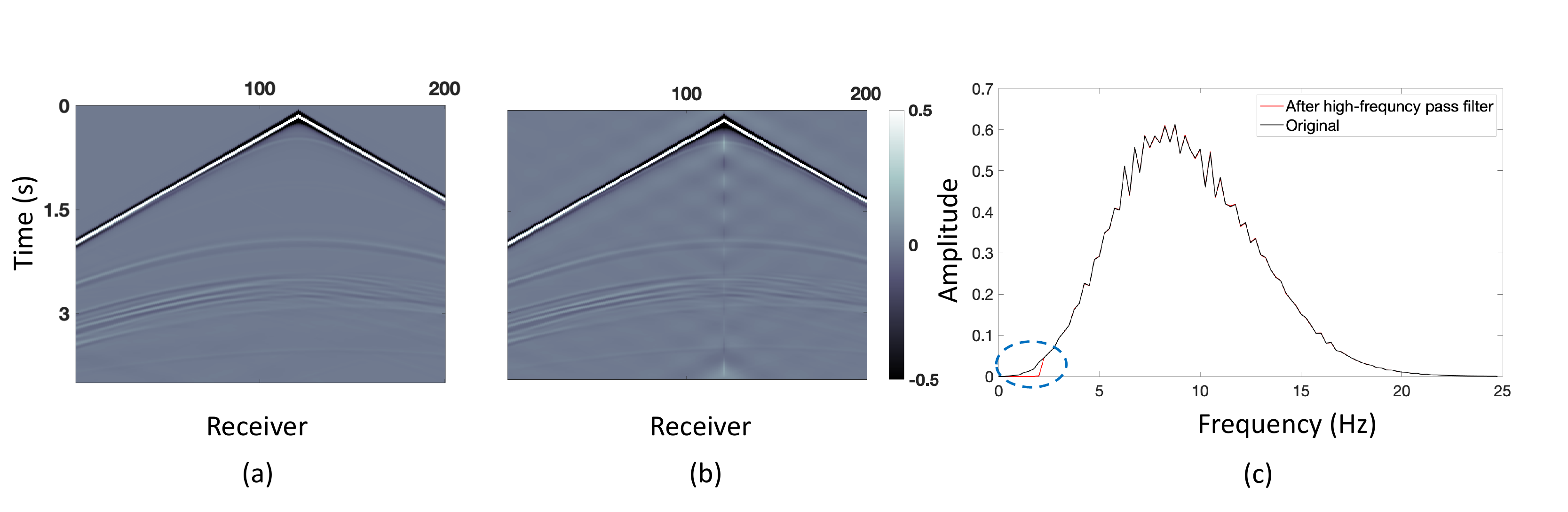}}
   \vskip -10pt
  \caption{Seismic shot gather for the testing velocity model in year 20: (a) without a high bandpass filter, and (b) with the Butterworth filter that has a cut-off frequency at 3~Hz. Plot (c) shows the frequency spectrum of the data. The blue dashed line on plot (c) marks the frequencies removed by the filter. } 
  \label{fig:hf-data}
\end{figure}

\begin{figure}
  \centering
  \subfigure{\includegraphics[trim=0cm 0cm 2cm 0cm,clip,scale=0.22,width=0.95\textwidth]{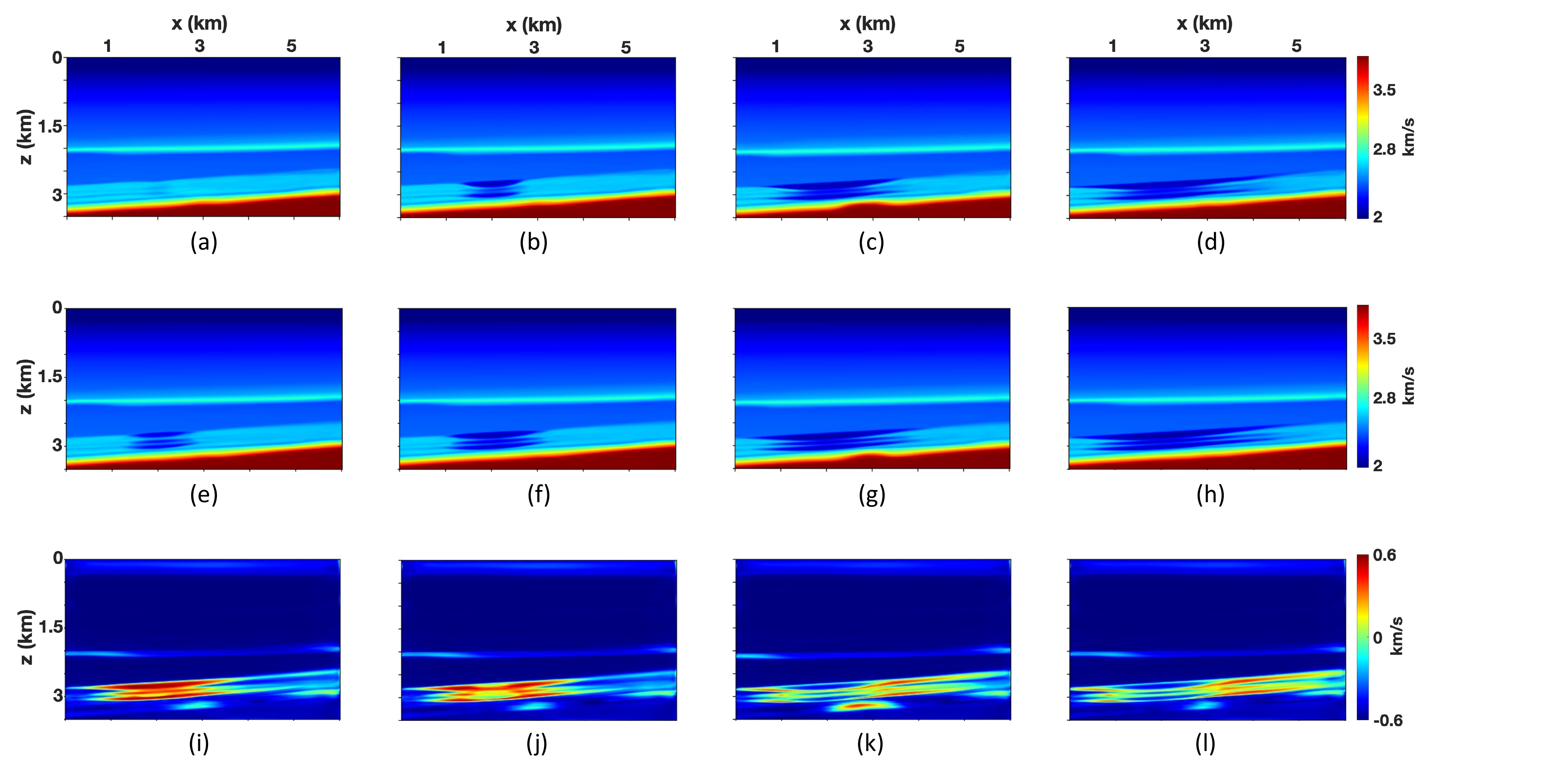}}
  \caption{P-wave velocity obtained from the testing seismic data using a high bandpass filter with a cut-off frequency of 3~Hz by (a-d) InversionNet and (e-h) InvNet${\_}$UQ. Plots (i-l) show the corresponding uncertainty from InvNet${\_}$UQ.} 
  \label{fig:hf}
\end{figure}

Compared with the benchmark results, the velocity models produced by InversionNet are noticeably distorted~(Figures~\ref{fig:hf}a-d), even though the low frequencies do not dramatically influence the input seismic data~(Figures~\ref{fig:hf-data}). Specifically, InversionNet fails to reconstruct small ${\mathrm{CO_2}}$ plumes, which could be indicative of ${\mathrm{CO_2}}$ leakages. InvNet\_UQ~(Figures~\ref{fig:hf}e-h) can still predict the plumes in the early stages of the injection with the resolution similar to that of the benchmark sections~(Figures~\ref{fig:clean}i-l), as revealed by the computed uncertainty map. The MSE also demonstrates that InvNet\_UQ~(\num{8.0e-4}) remains robust compared to the perfect data and achieves higher resolution than InversionNet~(\num{9.3e-4}). 

\subsection{Test of uncertainty-guided data augmentation}
\begin{figure}
  \centering
  \subfigure{\includegraphics[trim=0cm 0cm 16cm 0cm,clip,scale=0.22,width=0.6\textwidth]{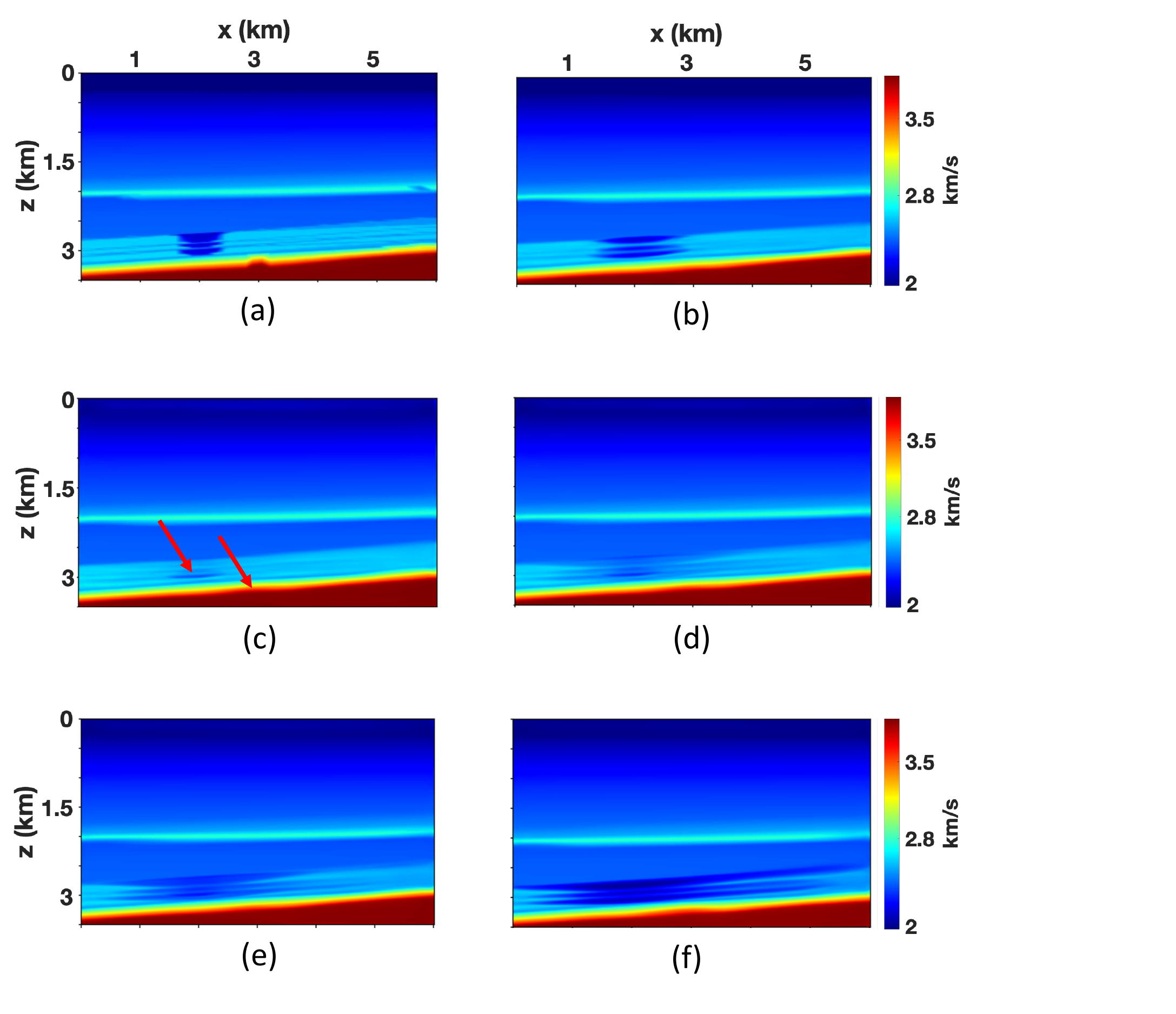}}
  \caption{Velocity model in year 1: (a) actual and (b) predicted. Plots (c) - (f) show the augmented data from the prediction.} 
  \label{fig:aug-data}
\end{figure}

We augment the velocity models predicted from the perfect data~(see equation~\ref{eq:Aug}). The uncertainty map is sampled with different weights, w=$[0.1, 0.3, 0.5, 0.7, 0.9]$. Compared to the prediction~(Figure~\ref{fig:aug-data}b), the augmented data~(Figure~\ref{fig:aug-data}c-f) contain small ${\mathrm{CO_2}}$ plumes and bumps in Zone~3~(marked by the red arrows), which improves the prediction of these features. These newly generated data are used together with the existing training data to retrain InvNet\_UQ. Then the trained network is applied to the testing data. 

\begin{figure}
  \centering
  \subfigure{\includegraphics[trim=0cm 3cm 2cm 0cm,clip,scale=0.22,width=0.95\textwidth]{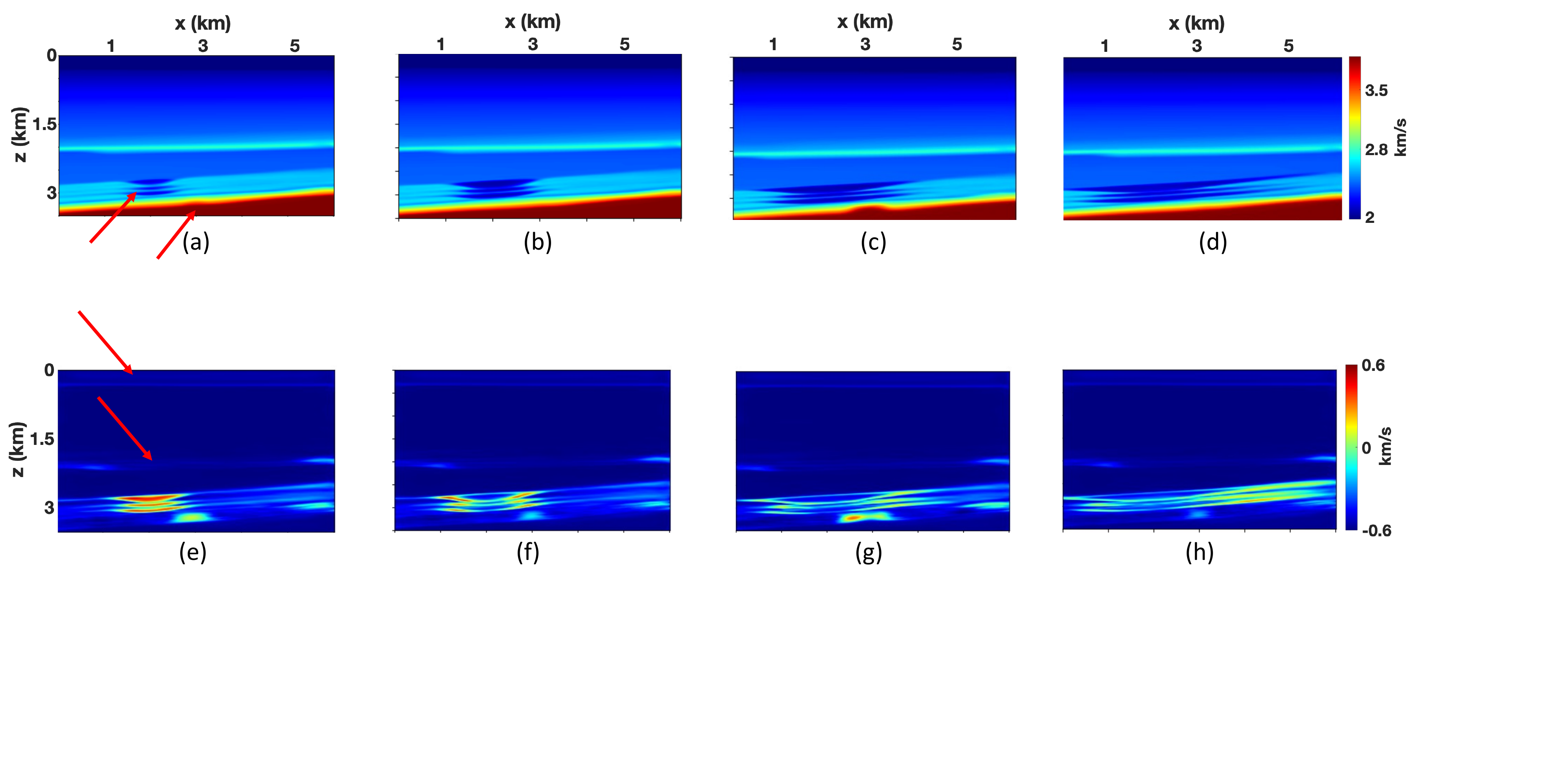}}
  \vskip -1cm
  \caption{P-wave velocity predicted by InvNet\_UQ trained using the augmented data in (a) year 1, (b) year 5, (c) year 20, and (d) year 150. Plots (e)-(h) show the corresponding uncertainty maps. } 
  \label{fig:clean-aug}
\end{figure}

The testing loss of the perfect data shows that training on the augmented data reduces the MSE of the testing data from \num{8.0e-4} to \num{6.0e-4}, which demonstrates the effectiveness of our data augmentation method. The detailed comparison between the velocity map~(Figures~\ref{fig:clean-aug}a-d) and the benchmark result~(Figures~\ref{fig:clean}i-l) shows that the network trained with the augmented data can improve the reconstruction of the velocity model. In particular, with the augmented data, the shape and amplitude of the predicted ${\mathrm{CO_2}}$ plumes are closer to those for the actual models, even in earlier years when the plumes are relatively small. In addition, the small bump in the high-velocity zone is captured more accurately. The calculated uncertainty further illustrates that the InvNet\_UQ has a much higher accuracy in predicting the ${\mathrm{CO_2}}$ plumes~(Zone 2), the low-velocity layer~(Zone 3), the high-velocity dipping layer~(Zone 1), and the upper part of the velocity model. Likewise, the ability of the proposed data-augmentation method to improve the prediction accuracy is demonstrated in the above tests for data with missing traces and missing low frequencies~(Table~\ref{tab:MSE}).

\section{Conclusions}
We developed a convolutional neural network~(CNN) with Simultaneous Quantile Regression~(SQR) to predict the time-lapse velocity model and estimate the uncertainty map from the input seismic data. SQR is designed to build a network that estimates the conditional distributions of all pertinent quantiles utilizing the pinball loss. Then the uncertainty map is obtained from the prediction interval. In contrast to most conventional UQ methods, the uncertainty calculated here is related to the predicted velocity and has a physical meaning~(i.e., the confidence interval). To fully utilize the computed uncertainty, we propose to employ data augmentation by sampling the uncertainty, adding the newly sampled data to the existing data set, and retraining the network. To benchmark the developed method, the pinball loss is replaced by the L1-norm objective function to form a deterministic neural network~(InversionNet).

We test InvNet${\_}$UQ on realistic synthetic data from the Kimberlina reservoir and compare the results with those from InversionNet. The results for clean data demonstrate that InvNet${\_}$UQ can predict the velocity model with sufficient accuracy. The estimated uncertainty map is consistent with the error map, which implies that the uncertainty can be used to evaluate the prediction accuracy. The tests on data with noise, missing traces, etc. show that InvNet${\_}$UQ remains robust and produces acceptable velocity maps, whereas the output of InversionNet is substantially distorted. This implies that the proposed method is applicable to field data and can facilitate confident decision-making. Retraining the network using the augmented data from the computed uncertainty improves the prediction accuracy, especially in areas with larger parameter variations including the ${\mathrm{CO_2}}$ plumes.




\section{Acknowledgments}
This work was funded by the U.S. Department of Energy~(DOE) Office of Fossil Energy’s Carbon Storage Research Program via the Science-Informed Machine Learning to Accelerate Real Time Decision Making for Carbon Storage~(SMART-CS) Initiative. Y. Liu and I. Tsvankin also acknowledge the support of the sponsors of the Center for Wave Phenomena~(CWP) in the Department of Geophysics at Colorado School of Mines.

\section{Author contributions statement}
Youzuo Lin proposed and supervised the research. Yanhua Liu and Xitong Zhang conceived the design of the experiment(s). Yanhua Liu conducted the experiment(s). All authors contributed ideas to the project, analyzed the results, and reviewed the manuscript. 



\bibstyle{seg}  
\bibliography{geophysics_endfloat}

\end{document}